\documentclass[
reprint,
superscriptaddress,
amsmath,
amssymb,
aps,
prl,
showkeys,
floatfix,
nobibnotes,
]{revtex4-1}

\usepackage{bm}
\usepackage{comment}
\usepackage{graphicx}
\usepackage[
colorlinks=true,
citecolor=myBlue,
linkcolor=BrickRed,
urlcolor=blue,
breaklinks=true,
]{hyperref}
\usepackage{mathtools}
\usepackage{multirow}
\usepackage{subfigure}
\usepackage[normalem]{ulem}
\usepackage{wasysym}
\usepackage[usenames,dvipsnames]{xcolor}
\usepackage[capitalize]{cleveref}
\usepackage{mathrsfs}
\usepackage{amsmath}
\usepackage{amssymb}
\usepackage{amsthm}
\usepackage{stmaryrd}

\makeatletter
\def\@fnsymbol#1{
	\ensuremath{\ifcase#1\or
	*\or				
	\ddagger\or			
	\dagger\or			
	\mathsection\or		
	\mathparagraph\or	
	\|\or				
	**\or				
	\ddagger\ddagger\or	
	\dagger\dagger		
	\else\@ctrerr\fi}}
\makeatother

\graphicspath{{figs/}}

\allowdisplaybreaks

\definecolor{myBlue}{rgb}{0.0, 0.47, 0.75}

\def\*#1{\mathbf{#1}}

\newcommand*\diff{\mathop{}\!\mathrm{d}}

\begin{document}

%
%

\title{Inherent--State Melting and the Onset of Glassy Dynamics \\ in Two-Dimensional Supercooled Liquids}

\author{Dimitrios Fraggedakis}
\thanks{Equal contributions}
\thanks{email: \href{mailto:dimfraged@gmail.com}{dimfraged@gmail.com}}

\affiliation{Department of Chemical \& Biomolecular Engineering, University of California, Berkeley, CA 94720}

\author{Muhammad R. Hasyim}
\thanks{Equal contributions}
\thanks{email: \href{mailto:muhammad\_hasyim@berkeley.edu}{muhammad\_hasyim@berkeley.edu}}

\affiliation{Department of Chemical \& Biomolecular Engineering, University of California, Berkeley, CA 94720}

\author{Kranthi K. Mandadapu}
\thanks{email: \href{kranthi@berkeley.edu}{kranthi@berkeley.edu}}

\affiliation{Department of Chemical \& Biomolecular Engineering, University of California, Berkeley, CA 94720}
\affiliation{Chemical Sciences Division, Lawrence Berkeley National Laboratory, Berkeley, CA 94720}

\date{\today}

%
%

\begin{abstract}
    Below the onset temperature $T_\text{o}$, the equilibrium relaxation time of most glass-forming liquids exhibits glassy dynamics characterized by super-Arrhenius temperature dependence. In this supercooled regime, the relaxation dynamics also proceeds through localized elastic excitations corresponding to hopping events between inherent states, i.e., potential-energy minimizing configurations of the liquid.
Despite its importance in distinguishing the supercooled regime from the high-temperature regime, the microscopic origin of $T_\text{o}$ is not yet known. 
Here, we construct a theory for the onset temperature in two dimensions and find that inherent-state melting transition, described by the binding-unbinding transition of dipolar elastic excitations, delineates the supercooled regime from the high-temperature regime. 
The corresponding melting transition temperature is in good agreement with the onset temperature found in various two-dimensional atomistic models of glass formers. 
We discuss the predictions of our theory on the displacement and density correlations of two-dimensional supercooled liquids, which are consistent with observations of the Mermin-Wagner fluctuations in recent experiments and molecular simulations.
\end{abstract}

\maketitle

%
%

\begin{figure*}[t]
    \centering
    \hspace{0.08in}\includegraphics[width=\textwidth]{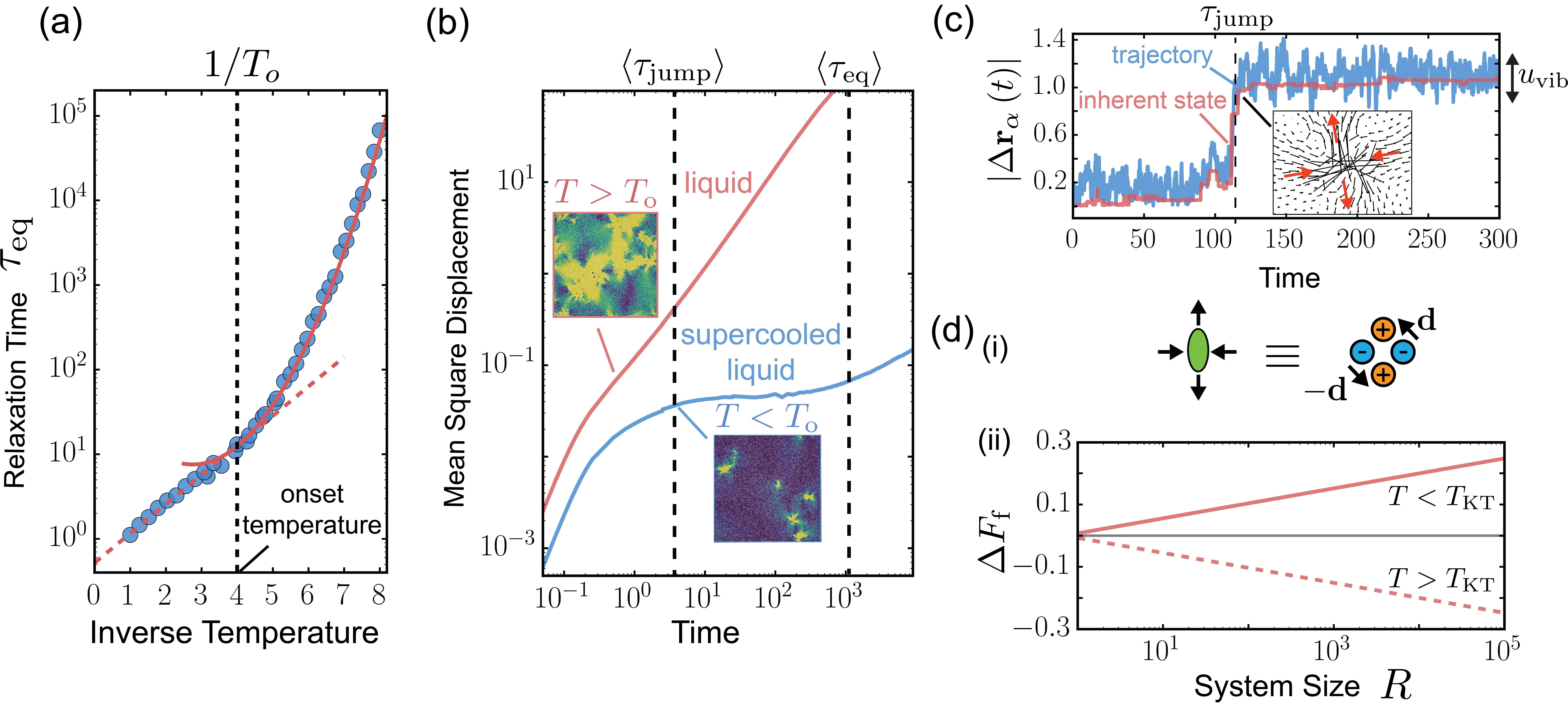}
    \caption{ 
    (a) Equilibrium relaxation time $\tau_\mathrm{eq}$ as a function of the inverse temperature. For $T>T_\mathrm{o}$, $\tau_\mathrm{eq}$ follows the classical Arrhenius behavior, however for $T<T_\mathrm{o}$, $\tau_\mathrm{eq}$ follows super-Arrhenius behavior.
    (b) Mean square displacement vs. time for $T>T_\mathrm{o}$ (red line) and $T<T_\mathrm{o}$ (blue line). Inset: Inherent-state (IS) particle displacement magnitude field showing the mobile regions at two different temperatures. 
    At $T<T_\mathrm{o}$, there exists an intermediate regime where only few localized mobile regions are observed. At $T>T_\mathrm{o}$, the system enters the diffusive regime immediately, with mobile regions spanning the entire system.  
    (c) A particle trajectory and its corresponding IS trajectory at $t=\tau_\mathrm{jump}$ when an excitation occurs. Inset: the corresponding IS displacement vector field showing the pure-shear deformation induced by the excitation. 
    (d)-(i) Correspondence of a pure-shear transformation, shown in inset of (c), with two bound elastic `dipoles' of net zero dipole moment. (d)-(ii) Free energy of formation, $\Delta F_\mathrm{f}$, of dipolar elastic excitations vs. system size $R$. For $T<T_\mathrm{KT}$, the formation of free dipoles is not energetically favorable, indicating the formation of bound dipoles at these temperatures instead, as in (d)-(i). For $T>T_\mathrm{KT}$, entropy changes the sign of $\Delta F_\mathrm{f}$ allowing the formation of free dipoles.
    In all cases, the parameters used correspond to the Poly-(12,0) model glass former (SM, Sec.~5.4). }\label{fig:figure_1}
\end{figure*}

\noindent\emph{Introduction}.--- The dynamics of glass-forming liquids slows down significantly below an onset temperature $T_\mathrm{o}$~\cite{Angell2000,cavagna2009supercooled,Elmatad2009,Katira2019}, as seen in the cross-over from Arrhenius ($T>T_\mathrm{o}$) to super-Arrhenius ($T<T_\mathrm{o}$) growth of the equilibrium relaxation time $\tau_\mathrm{eq}$; see \cref{fig:figure_1}(a).
The cross-over is also observed in the mean square displacement (MSD)~\cite{berthier2011theoretical}, as shown in \cref{fig:figure_1}(b). 
Above $T_\mathrm{o}$, MSD is characterized only by the ballistic and diffusive regimes~\cite{zwanzig2001nonequilibrium}. 
For $T<T_\mathrm{o}$, however, a new intermediate (glassy) regime appears where the MSD exhibits a plateau-like shape that is reminiscent of solids~\cite{vineyard1958scattering,schroder2020solid}. 
In two dimensions (2D) specifically, this solid-like behavior manifests as Mermin-Wagner fluctuations \cite{mermin1966absence,mermin1968crystalline}, which are long-wavelength fluctuations typically associated with 2D elastic solids. It has been shown in recent experiments involving 2D colloidal systems as well as molecular simulations that such fluctuations affect the finite-size scaling of the MSD and density autocorrelations~\cite{flenner2015fundamental,shiba2016unveiling,illing2017mermin,vivek2017long,tarjus2017glass}. 

The supercooled liquid, i.e., liquid below $T_\mathrm{o}$, is further characterized by dynamical heterogeneity \cite{berthier2011dynamical}, where particles initially move in sparse `mobile' regions that spread over time~\cite{Keys2011,guiselin2022microscopic}.
These initial mobile regions are spatially localized and can be classified as excitations~\cite{Keys2011,hasyim2021theory} that drive particle hopping dynamics, while the rest of the system vibrates around its initial state (inset of \cref{fig:figure_1}(b)). 
The excitations correspond to hopping events between neighboring inherent-state (IS) configurations, which are the energy-minimized configurations in the potential energy landscape \cite{stillinger1982hidden}.
This excitation-based perspective is central to the dynamical facilitation (DF) theory \cite{chandler2010dynamics}, one of the theories describing the super-Arrhenius relaxation behavior, where dynamical heterogeneity is understood in terms of excitations facilitating the formation and relaxation of nearby excitations in a hierarchical manner \cite{Garrahan2002,Keys2011}.
For $T>T_\mathrm{o}$, however, particles may move more diffusively, suggesting a different relaxation mechanism where motion occurs with little to no dynamical heterogeneity (inset of \cref{fig:figure_1}(b)). 

While various theories \cite{chandler2010dynamics,reichman2005mode,biroli2012random,Dyre2006} have attempted to explain the dynamics of liquids below the onset temperature $T_\mathrm{o}$, none have so far identified the physical nature of this crossover, thereby determining the value of $T_\mathrm{o}$ itself.
Taken together, the qualitative differences on the dynamics above and below $T_\mathrm{o}$ raise the following questions:
(i) What is the microscopic origin of $T_\mathrm{o}$ that distinguishes normal liquids from supercooled liquids? (ii) Does $T_\mathrm{o}$ signal a change in the relaxation mechanism between these two regimes? And, (iii) how is $T_\mathrm{o}$ connected to the solid-like nature of the supercooled liquid at intermediate timescales?

In this work, we address these questions related to 2D supercooled liquids by constructing a theory where the origin of $T_\mathrm{o}$ lies within the statistical mechanics of excitation events.
To demonstrate this, we focus on the time evolution of particles within the intermediate timescales of
\begin{equation}
\tau_\mathrm{vib}(T)  \ll t \sim \langle \tau_\mathrm{jump}(T) \rangle \ll \tau_\mathrm{eq}(T) \,, \label{eq:inttimescale}
\end{equation}
where $\tau_\mathrm{vib}(T)$ is the characteristic vibrational timescale and $\langle \tau_\mathrm{jump}(T) \rangle$ corresponds to the average time needed for a particle to hop to its next position; see \cref{fig:figure_1}(c).
The particle dynamics at these timescales are characterized by the instantaneous IS positions $\* R^\alpha(t)$. 
The IS trajectory coarse-grains the vibrational motion, and hopping in particle motions are reflected as jumps in the IS positions at $t=\tau_\mathrm{jump}$~\cite{Schroder2000}. 
Within the perspective of DF theory \cite{chandler2010dynamics,Keys2011}, recent work \cite{hasyim2021theory} indicates that jumps in the inherent states correspond to excitation events, which induce localized pure-shear deformation \cite{hasyim2021theory,chacko2021elastoplasticity} (inset of \cref{fig:figure_1}(c)). 
One of the main ideas of the present work is that, analogous to electrostatics, an excitation in 2D can be modeled as two bound elastic `dipoles'~\cite{moshe2015geometry,moshe2015elastic}; see \cref{fig:figure_1}(d)-(i) and Sec.~2.2 of the Supplemental Material (SM). 
If dipoles are considered to be the fundamental units of excitations, then an energy-entropy argument for their formation hints towards a transition temperature $T_\mathrm{KT}$ (\cref{fig:figure_1}(d)-(ii) and SM, Sec.~3.1) that governs the unbinding of localized excitations into free dipoles, and thus leads to a change in relaxation mechanism. 
Furthermore, the binding-unbinding transition is similar to the one described by the Kosterlitz-Thouless-Halperin-Nelson-Young (KTHNY) theory~\cite{kosterlitz1972long,kosterlitz1973ordering,halperin1978theory,nelson1978study,nelson1979dislocation,young1979melting} of dislocation-mediated melting. 
This, in turn, provides an alternative picture of the transition in terms of inherent states `melting' into a high-temperature fluid, with the transition temperature being the onset temperature. 

\begin{figure}[t]
    \centering
    \hspace{0.08in}\includegraphics[width=0.45\textwidth]{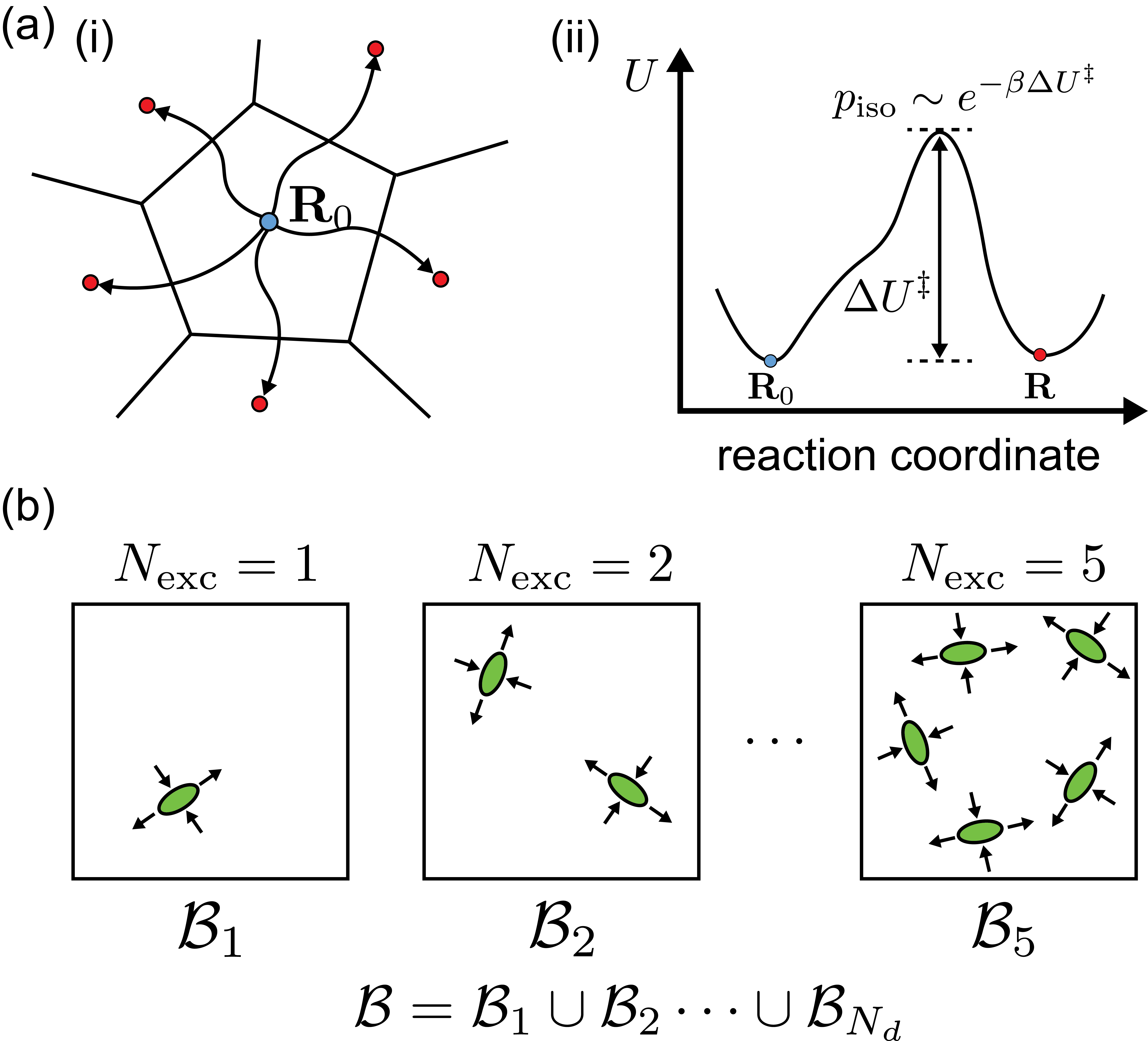}
    \caption{(a)-(i) Illustration of inherent-state (IS) jumps in configuration space obtained using the isoconfigurational ensemble. (a)-(ii) Potential energy landscape that defines the conditional probability, $p_\mathrm{iso}$, for visiting state $\mathbf{R}$ starting from $\mathbf{R}_0$ in terms of the transition state energy barrier $\Delta U^\ddag$. (b) Schematic of the subsets of the set of all possible nearest-neighboring inherent states $\mathcal{B}$, organized in terms of the excitation number. In this schematic, excitations lead to localized pure-shear transformations in the medium, consistent with \cite{hasyim2021theory,chacko2021elastoplasticity}.}\label{fig:figure_2} 
\end{figure}

\noindent\emph{Theory}.--- To establish a thermodynamic framework for understanding the onset temperature $T_\mathrm{o}$ via excitation events, we begin by constructing an isoconfigurational ensemble~\cite{widmer2004reproducible}, corresponding to jumps between a given IS and its neighboring states in the potential energy landscape.
Conceptually, such a construction can be understood by considering an IS trajectory in \cref{fig:figure_1}(c), where
a single trajectory provides one sample realization of a single jump that takes a system from the initial IS configuration $\{ \mathbf{R}_0^\alpha \}$ to one of its neighboring IS configurations at $t=\tau_\mathrm{jump}$. 
The isoconfigurational ensemble can then be built by initiating multiple trajectories from $\{ \mathbf{R}_0^\alpha \}$, which eventually visit all possible neighboring ISs (\cref{fig:figure_2}(a)-(i)). 
The set of all neighboring ISs of $\{ \* R_0^\alpha \}$, $\mathcal{B}(\{ \* R_0\} )$, sampled by the isoconfigurational ensemble, forms a basis for an ensemble of excitation configurations.
Given the energy-landscape complexity, not all neighboring ISs are visited with the same frequency. Under the assumptions of transition-state theory~\cite{chandler1978statistical,peters2017reaction}, we can compute the conditional probability to visit a neighboring IS as $p_\mathrm{iso} \sim e^{-\beta \Delta U^\ddag}$ (\cref{fig:figure_2}(a)-(ii) and SM, Sec.~1.1), where $\Delta U^\ddag$ is the potential energy difference between initial and transition state, while $\beta=1/k_\mathrm{B}T$ and $k_\mathrm{B}$ is the Boltzmann constant. The normalization factor of $p_\mathrm{iso}$ defines the isoconfigurational partition function $Q_\mathrm{iso}(\{ \* R_0^\alpha \})=\sum_{\{ \mathbf{R}^\alpha \} \in \mathcal{B}(\{ \* R_0^\alpha\})} e^{-\beta \Delta U^\ddag}$, and can be used to study the statistics of IS jumps from all possible initial ISs by averaging it over the IS ensemble, leading to an IS-averaged partition function (SM, Secs.~1.1 and 1.2)
\begin{equation}\label{eq:Q_bar_iso}
    \bar{Q}_\mathrm{iso} = \left\langle\sum_{\{ \mathbf{R}^\alpha \} \in \mathcal{B}(\{ \* R_0^\alpha\})} e^{-\beta \Delta U^\ddag}\right\rangle_\mathrm{IS} \,,
\end{equation}
where $\bar{Q}_\mathrm{iso} = \langle Q_\mathrm{iso}(\{ \* R_0^\alpha \}) \rangle_\mathrm{IS}$ and $\left\langle \ldots \right\rangle_\mathrm{IS}$ is an ensemble average over all possible ISs. 

An IS jump can lead to the formation of multiple excitations in space. We can thus express $\mathcal{B}$ as a union of subsets that contain $N_\mathrm{exc}$ excitations,
$\mathcal{B}=\bigcup_{N_\mathrm{exc}=1}^{N_\mathrm{d}} \mathcal{B}_{N_\mathrm{exc}}$, 
where $N_\mathrm{d}$ is the maximum number of excitations bounded by the total number of particles (\cref{fig:figure_2}(b)). 
This representation allows us to consider excitations as quasi-particles described by their positions $\{ \mathbf{q}^\mu\}$ and internal degrees of freedom $\{\mathbf{s}^\mu \}$. 
As a result, we derive from \cref{eq:Q_bar_iso} a grand canonical partition function for a thermodynamic ensemble of excitations (SM, Sec.~1.2),
\begin{subequations}
    \begin{alignat}{2}
        \bar{\Xi}_\mathrm{exc} & = 1 + \bar{Q}_\mathrm{iso} = \sum_{N_\mathrm{exc}=0}^{N_\mathrm{d}} Z_{N_\mathrm{exc}} \tilde{y}^{N_\mathrm{exc}} \label{eq:Ksi_bar_iso} \,, \\
        Z_{N_\mathrm{exc}} & = \frac{1}{ N_\mathrm{exc}!} \int \prod_{\mu=1}^{N_\mathrm{exc}} \frac{\diff^d \mathbf{q}^\mu \diff^{d_\mathrm{int}} \mathbf{s}^\mu}{(a_\mathrm{exc})^{d N_\mathrm{exc}}}  e^{-\tilde{\mathcal{H}}}\,, \label{eq:Z_part_fun}
    \end{alignat}
\end{subequations}
where $a_\mathrm{exc}$ is the size of excitations, $\tilde{y} = e^{-\tilde{E}_\mathrm{c}}$ is a fugacity parameter controlling the concentration of excitations via its dimensionless self energy $\tilde{E}_\mathrm{c}=\beta\bar{E}_\mathrm{c}$, while $\tilde{\mathcal{H}}=\sum_{\langle \mu, \gamma \rangle} \tilde{v}^{\mu \gamma}$, with $\tilde{v}^{\mu \gamma}=\beta\bar{v}^{\mu \gamma}$, is the dimensionless total interaction energy of pairs of excitations. 

It is notable that through \cref{eq:Ksi_bar_iso}, we can find the equilibrium concentration of non-interacting excitations as $c_\mathrm{eq}(T) \sim e^{-\beta \bar{E}_\mathrm{c}}$ (SM, Sec.~1.3). This Arrhenius form of $c_\mathrm{eq}(T)$ is found in kinetically constrained models \cite{Garrahan2003,Ritort2003}, used by DF theory to model hierarchical relaxation between excitations \cite{chandler2010dynamics}. 
It is also consistent with the rate of particle-hopping events $c_\sigma(T)$, which is a proxy for $c_\mathrm{eq}(T)$ in molecular simulations for DF theory, since $c_\sigma(T)$ is empirically observed to be of Arrhenius form \cite{Keys2011, hasyim2021theory}.

The framework of geometric charges~\cite{moshe2015geometry,moshe2015elastic} allows us to describe the formation of excitations in glass formers. 
Geometric monopoles, dipoles, quadrupoles or higher order multipoles are therefore candidates for describing elementary excitations 
\footnote{See also Sec.~2.2 of the Supplementary Material for a discussion of how geometric charges can be mapped to familiar defects in crystalline solids, such as disclinations, dislocations, and point defects.}. 
To determine which geometric charges are thermodynamically admissible, we resort to their free energy of formation in an elastic medium, $\Delta F_\mathrm{f}$ (SM, Sec.~3.1). 
For monopoles with charge $m$, we find $\Delta F^\mathrm{mnpl}_\mathrm{f} \sim m R^2 \ge 0$ for all $R$ and $T$, where $R$ is the size of the system. 
We therefore conclude that geometric monopoles are not thermodynamically favorable. For dipoles with dipole-moment magnitude $d_\mathrm{c}$, the free energy is $\Delta F^\mathrm{dpl}_\mathrm{f} \sim \left(\frac{d_\mathrm{c}^2 Y }{8\pi} - 2k_\mathrm{B} T\right)\ln R$, where $Y$ is the Young's modulus; see \cref{fig:figure_1}(d)-(ii). When $k_\mathrm{B} T < d_\mathrm{c}^2 Y/16\pi$, we find $\Delta F^\mathrm{dpl} > 0$ which implies that spontaneous formation of single dipoles is not favored. 
For $k_\mathrm{B} T > d_\mathrm{c}^2 Y/16\pi$, however, the free energy becomes negative, $\Delta F^\mathrm{dpl}_\mathrm{f} < 0$, and free dipole formation is preferred. 
This qualitative change on the sign of $\Delta F^\mathrm{dpl}_\mathrm{f}$ at $T_\mathrm{c} = d_\mathrm{c}^2 Y/16\pi k_\mathrm{B}$ hints towards a binding-unbinding transition similar to the KTHNY theory. 
For quadrupoles, $\Delta F^\mathrm{qdpl}_\mathrm{f} \sim -\ln R$ which leads to $\Delta F^\mathrm{qdpl}_\mathrm{f} < 0$ for large $R$, and thus formation of quadrupoles is always thermodynamically admissible for all $T$. 
Quadrupoles consist of two bound dipoles in the limit of infinitesimal separation (\cref{fig:figure_1}(d)-(i)), and this motivates us to investigate the binding-unbinding transition through similar free energy arguments. In particular, we find the temperature for which two free dipoles are preferred when compared to the bound dipole-pair state to be $T_\mathrm{c}^\prime=d_\mathrm{c}^2 Y/8\pi k_\mathrm{B}$ (SM, Sec.~3.1). 
The observation that this transition temperature differs from that of the free-energy argument of a single dipole is a result of neglecting interactions between the two dipoles. Thus, a more complete picture that involves a grand canonical ensemble of many interacting dipolar excitations is required to understand the binding-unbinding transition, which we proceed to analyze.

An ensemble of interacting dipoles is described by their (dimensionless) self energy $\tilde{E}_\mathrm{c}$ and interaction $\tilde{v}^{\mu \gamma}$, which can be derived following the use of charge and dipole-moment conservation laws (SM, Secs.~2.3~and~2.4). This yields,  
\begin{subequations}
    \begin{alignat}{2}
        \tilde{E}_\mathrm{c} & =  \frac{\tilde{Y}^\mathrm{IS}}{8\pi}  (\tilde{C}+1) \,, \label{eq:self_energy} \\
        \tilde{v}^{\mu \gamma} & = 
        \frac{\tilde{Y}^\mathrm{IS}}{4\pi} \left(\tilde{d}_i^\mu \tilde{d}_i^\gamma \left(1-\ln\frac{q^{\mu \gamma}}{a_\mathrm{dpl}}\right) -\frac{\tilde{d}_i^\mu q_i^{\mu\gamma} \tilde{d}_j^\gamma q_j^{\mu\gamma}}{\left(q^{\mu\gamma}\right)^2}\right) \,, \label{eq:interaction_energy}
    \end{alignat}
\end{subequations}
where $\tilde{Y}^\mathrm{IS} = \beta d_\mathrm{c}^2 Y^\mathrm{IS}$ with $Y^\mathrm{IS}$ being the IS Young's modulus, $q_i^{\mu\gamma} = q_i^\mu - q^\gamma_i$, $a_\mathrm{dpl}$ is the dipole size, $\tilde{d}_i^\mu = d_i^\mu/d_\mathrm{c}$ is the dimensionless dipole-moment vector, and the dipole moment magnitude $d_\mathrm{c}$ and $\tilde{C}$ are as yet undetermined constants; see Secs.~3.2~and~3.4 of the SM for the derivation of \cref{eq:self_energy,eq:interaction_energy}. We determine $\tilde{C}$ and $d_\mathrm{c}$ following the mapping between bound geometric dipoles and localized pure-shear excitations (SM, Sec.~3.3), which were modeled previously via a force-dipole formalism \cite{hasyim2021theory}. 
This mapping ensures equivalency between the average energy barrier and the spatial stress distributions corresponding to excitations, and yields (SM, Sec.~3.3)
\begin{equation}
d_\mathrm{c} = \frac{2 \pi R_\mathrm{exc} \epsilon_\mathrm{c}}{\nu^\mathrm{IS}+1}, \quad \tilde{C} = \frac{3+\nu^\mathrm{IS}}{4}, 
\end{equation}
where $R_\mathrm{exc} = a_\mathrm{dpl}/\sqrt{2}$, and the eigenstrain, $\epsilon_\mathrm{c}$, are determined from the knowledge of local structure; see Eqs.~(34)-(35) in~\cite{hasyim2021theory}.

Similar to the KTHNY theory \cite{kosterlitz1972long,kosterlitz1973ordering,halperin1978theory,nelson1978study,nelson1979dislocation,young1979melting}, we now study the dipole binding-unbinding transition via its impact on the elastic response of an IS in the presence of excitations. 
The dimensionless stiffness tensor $\tilde{C}_{ijkl}^\mathrm{R}$ governing this response can be written using any combination of two elastic constants due to isotropy of glass formers. 
Choosing the Young's ($Y^\mathrm{R}$), and shear ($G^\mathrm{R}$) moduli for convenience, a static linear-response theory yields (SM, Secs.~4.1 and 4.2)
\begin{subequations}
    \begin{alignat}{2}
        \frac{1}{\tilde{G}^\mathrm{R}} &= \frac{1}{\tilde{G}^\mathrm{IS}} +\frac{A_0}{d_\mathrm{c}^2}\left( \langle \hat{\epsilon}_{ij}^\mathrm{e} \hat{\epsilon}_{ij}^\mathrm{e} \rangle -\frac{1}{2}\langle \hat{\epsilon}_{ii}^\mathrm{e} \hat{\epsilon}_{kk}^\mathrm{e} \rangle \right) \,, \label{eq:shearrenorm}
        \\
        \frac{1}{\tilde{Y}^\mathrm{R}} &= \frac{1}{\tilde{Y}^\mathrm{IS}} +\frac{A_0}{4d_\mathrm{c}^2} \left( \langle \hat{\epsilon}_{ij}^\mathrm{e} \hat{\epsilon}_{ij}^\mathrm{e} \rangle +\frac{1}{2}\langle \hat{\epsilon}_{ii}^\mathrm{e} \hat{\epsilon}_{kk}^\mathrm{e} \rangle \right) \,, \label{eq:youngrenorm}
    \end{alignat}
\end{subequations}
where $A_0$ is the area of the medium, $\hat{\epsilon}^\mathrm{e}_{ij}$ is the area-averaged elastic strain due to presence of geometric dipoles, and $\langle \ldots \rangle$ denotes the grand-canonical ensemble average.
The transition is then determined by locating the point above which $\tilde{G}^\mathrm{R} = 0$ and $\tilde{Y}^\mathrm{R}=0$, corresponding to the loss of elastic moduli signalling a dipole-mediated melting of an IS.

Since the fugacity $\tilde{y}$ is small Table~1 of SM, Sec~5.1), \cref{eq:shearrenorm,eq:youngrenorm} can be evaluated by a fugacity series expansion around $\tilde{y}=0$; such a procedure, however, leads to divergent expressions near the unbinding/melting transition.
This situation can be remedied via the renormalization group (RG) procedure~\cite{Goldenfeld2018,nelson1979dislocation,kardar2007statistical}, which uses the initial fugacity expansion to obtain the following set of RG equations for the fugacity, Young's and shear moduli (SM, Secs.~5.1~and~5.2):
\begin{subequations}
    \begin{alignat}{3}
    \frac{\diff \tilde{y}}{\diff \ell}  
        & = \left(2-\frac{\tilde{Y}}{8 \pi}\right) \tilde{y}+ 2 \pi \tilde{y}^2 e^{\frac{\tilde{Y}}{16\pi}}  I_0\left(\tfrac{\tilde{Y}}{8 \pi}\right) \label{eq:rgflowfugacity} \,,
        \\
        \frac{\diff \tilde{Y}^{-1}}{\diff \ell}  
        & = \frac{\pi^2}{4} \tilde{y}^2 e^{\frac{\tilde{Y}}{8 \pi}} \left(2I_0 \left(\tfrac{\tilde{Y}}{8 \pi }\right)-I_1 \left(\tfrac{\tilde{Y}}{8 \pi }\right)\right) \label{eq:rgflowyoung} \,, \\
        \frac{\diff \tilde{G}^{-1}}{\diff \ell} 
        & = \pi^2 \tilde{y}^2 e^{\frac{\tilde{Y}}{8 \pi}} I_0 \left(\tfrac{\tilde{Y}}{8 \pi }\right) \label{eq:rgflowshear} \,,
    \end{alignat}
\end{subequations}
where $I_n(x)$ is the $n$-th order modified Bessel function of the first kind and $\ell$ is associated with the logarithm of a lengthscale. 
The renormalized elastic moduli are then obtained by integrating \cref{eq:rgflowfugacity,eq:rgflowyoung,eq:rgflowshear} to the large-$\ell$ limit with initial conditions given by $\tilde{G}(0) = \tilde{G}^\mathrm{IS}$, $\tilde{Y}(0) = \tilde{Y}^\mathrm{IS}$, and $\tilde{y}(0) = e^{-\tilde{E}_\mathrm{c}}$ where $\tilde{E}_\mathrm{c}$ is provided by \cref{eq:self_energy}.

\Cref{eq:rgflowfugacity,eq:rgflowshear,eq:rgflowyoung} are very similar to the RG equations of the KTHNY theory ~\cite{nelson1979dislocation,young1979melting} with the exception of the $\pi^2$ factor in \cref{eq:rgflowshear,eq:rgflowyoung} replacing the usual factor of $3 \pi$. 
This difference stems from the continuous orientability of dipoles in an amorphous medium, in contrast to crystalline solids where only discrete values are allowed. 
Despite this, the difference yields only a minor change in the critical exponent describing the vanishing elastic moduli near $T_\mathrm{KT}$ (SM, Sec.~5.3).
Nevertheless, from the RG flow equations we observe that working only with $\tilde{Y}$ and $\tilde{y}$ is sufficient to understand the melting transition, given all flow equations depend exclusively on these two variables. 

\begin{figure}[t]
    \centering
    \hspace{0.08in}\includegraphics[width=0.415\textwidth]{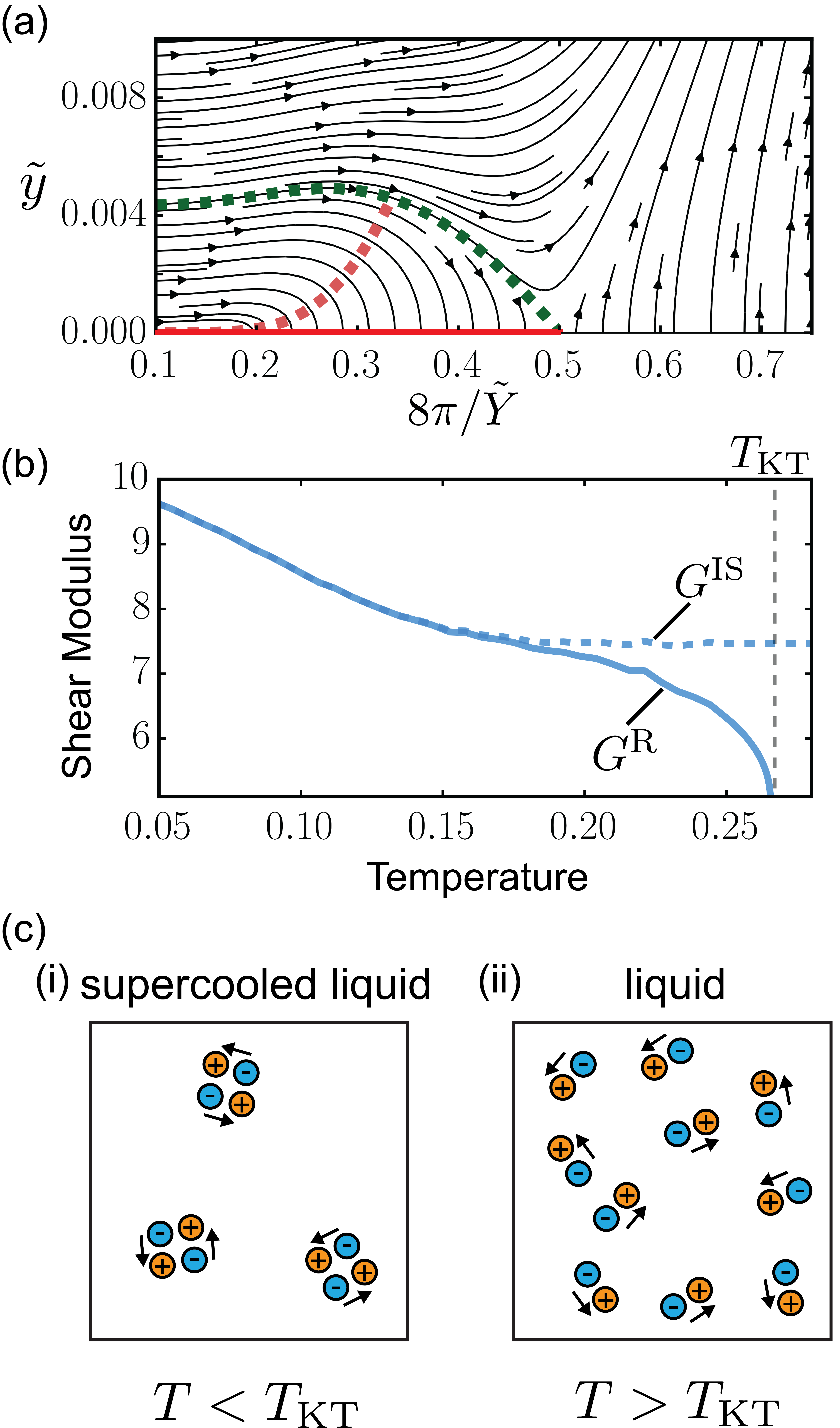}
    \caption{(a) Phase-space portrait of the RG equations in \cref{eq:rgflowyoung,eq:rgflowfugacity}. Horizontal red line corresponds to the locus of fixed points terminating at $\tilde{Y} = 16 \pi$. The separatrix (green dashed line), distinguishes the supercooled regime from the high-temperature regime. Red dashed line is the possible initial conditions for the RG flow. (b) Bare, $G^\mathrm{IS}$, and renormalized, $G^\mathrm{R}$, shear modulus vs. temperature. As $T$ is increased, $G^\mathrm{IS}$ shows a linear dependence in $T$ until a temperature where $G^\mathrm{IS}$ reaches a plateau. Meanwhile, $G^\mathrm{R}$ moves away from $G^\mathrm{IS}$ as $T$ is increased and drops to zero at $T=T_\mathrm{KT}$, where the supercooled liquid phase loses its rigidity. In (a) and (b) the parameters used correspond to the Poly-(12,0) model. (c) Schematic representation on the different mechanism of relaxation between (i) supercooled liquids and (ii) liquids. Supercooled liquids relax by forming bound dipolar elastic excitations, while at high temperatures, they relax through free dipolar excitations. }\label{fig:figure_3}
\end{figure}

$\\$\emph{Results}.--- We start by analyzing \cref{eq:rgflowfugacity,eq:rgflowyoung} in terms of their fixed points $(\tilde{y}^*,\tilde{Y}^*)$, where we find that for $\tilde{y}^*=0$ and any value of $\tilde{Y}^*$ the RG flow equations are stationary (SM, Sec.~5.3). 
This behavior is seen in \cref{fig:figure_3}(a), where we show the phase portrait for $\tilde{y}$ and $8\pi/\tilde{Y}$. 
We observe that for any initial point starting within the region below the separatrix (green-dashed line), the flow converges towards the locus of $\tilde{y}^*=0$ and $\tilde{Y}^* \neq 0$ (red line), indicating the existence of a solid phase where a supercooled liquid behaves elastically at the intermediate timescales. 
For $\tilde{Y}^* \le 16\pi$, however, the family of fixed points becomes unstable to any infinitesimal perturbations around $\tilde{y}^*=0$ (SM, Sec.~5.3), indicating the fluid phase. This implies that the separatrix controls the location of the melting point. 
Thus, the melting temperature $T_\mathrm{KT}$ is obtained by finding the initial conditions $(\tilde{Y}^\mathrm{IS}(T_\mathrm{KT}),\tilde{y}(T_\mathrm{KT}))$ that lie on the separatrix so that the RG flow converges to the fixed point $(\tilde{Y}^\mathrm{R}, \tilde{y}_\mathrm{R}) =( 16\pi,0)$. 
In \cref{fig:figure_3}(a), we show, in red dashed line, a curve of initial conditions that terminates at the separatrix.

To validate our hypothesis that $T_\mathrm{KT}$ corresponds to the onset temperature for glassy dynamics $T_\mathrm{o}$, we test the theory on six models of glass-forming liquids (SM, Sec.~5.4). 
Here, we take the perspective of DF theory for estimating $T_\mathrm{o}$, which is done by fitting the parabolic law, $\ln \tau_\mathrm{eq} \sim J^2(\beta- \beta_\mathrm{o})^2$, to the relaxation-time data \cite{Elmatad2009,Katira2019}, where $\beta_\mathrm{o} = 1/k_\mathrm{B} T_\mathrm{o}$ and $J$ is an effective energy scale. 
Estimation of $T_\mathrm{o}$ can also be done via computation of the particle-hopping rate $c_\sigma(T)$ from coarse-grained particle trajectories, which is empirically observed to be Arrhenius at $T \le T_\mathrm{o}$ at short intermediate timescales, i.e., $c_\sigma(T) \sim e^{-J_\sigma(\beta -\beta_\mathrm{o})}$ with $J_\sigma \sim J$, indicating the onset of activated dynamics \cite{Keys2011,hasyim2021theory}. 
Both estimates have been shown to agree with each other \cite{Keys2011,hasyim2021theory}, and we choose the parabolic-law fitting for this work. 
The IS melting transition is evaluated in two ways: (1) the approximate estimate $T_\mathrm{KT}^\mathrm{app} = d_\mathrm{c}^2 Y^\mathrm{IS}/16\pi k_\mathrm{B}$ that assumes $\tilde{y} = 0$, and thus no renormalization occurs on the Young's modulus, and (2) the true estimate $T_\mathrm{KT}$ based on the intersection of the separatrix with the curve of initial conditions for different models, where numerical integration of  \cref{eq:rgflowfugacity,eq:rgflowyoung} is performed (SM, Sec.~5.4). 

\Cref{tab:predictions_KT} summarizes the results of our theory compared to the estimated $T_\mathrm{o}$ for these models. In all cases, the true estimate of the melting temperature is in reasonable agreement with the observed onset temperature. 
On the contrary, $T_\mathrm{KT}^\mathrm{app}$ typically overestimates the transition point, which may be attributed to ignoring the renormalization of the Young's modulus. In \cref{fig:figure_3}(b), we also plot $G^\mathrm{IS}$ and $G^\mathrm{R}$ as a function of temperature, where we see that $G^\mathrm{R} < G^\mathrm{IS}$ due to a softening effect of the excitations on the elastic stiffness of the inherent states. This observation also extends to the Young's modulus, i.e., $Y^\mathrm{R} < Y^\mathrm{IS}$ (Fig.~S.21 in SM, Sec.~5.4).

\begin{table}[t]
\begin{ruledtabular}
\begin{tabular}{lccc}
Model  & $T_\mathrm{o}$ & $T_\mathrm{KT}^\mathrm{app}$ & $T_\mathrm{KT}$\\ \hline
Poly-(12,0), ($\varepsilon=0.2$)    & 0.25 & 0.38 & 0.27 \\
Poly-(12,6), ($\varepsilon=0.2$)    & 0.17 & 0.16 & 0.11 \\
Poly-(18,0), ($\varepsilon=0.0$)    & 1.10 & 2.00 & 1.40 \\
Poly-(18,0), ($\varepsilon=0.2$)    & 0.39 & 0.51 & 0.35 \\
Poly-(10,6), ($\varepsilon=0.1$)    & 0.17 & 0.35 & 0.24 \\
Poly-(10,6), ($\varepsilon=0.2$)    & 0.14 & 0.15 & 0.10 \\
\end{tabular}
\end{ruledtabular}
\caption{Comparison of the onset temperature $T_\mathrm{o}$, obtained through parabolic-law fitting of $\tau_\mathrm{eq}$ \cite{Elmatad2009,Katira2019}, with the predicted transition temperature $T_\mathrm{KT}$ for the binding-unbinding transition of dipolar elastic excitations for six model glass-forming liquids. The predicted $T_\mathrm{KT}$ is obtained by integrating the RG flow equations \cref{eq:rgflowshear,eq:rgflowyoung}. For completeness, we also report the approximated transition temperatures $T_\mathrm{KT}^\mathrm{app}$, which assumes that $\tilde{Y}^\mathrm{R} \approx \tilde{Y}^\mathrm{IS}$. For further numerical details, see SM, Sec.~5.4.} \label{tab:predictions_KT}
\end{table}

Another way to interpret these results is in terms of the relaxation mechanism of glass-forming liquids. 
The RG analysis indicates that bound dipolar elastic excitations are favored within the supercooled regime. 
Since our theory governs the nature of IS jumps, we find that below $T_\mathrm{o}$ localized mobile regions are intimately linked to bound dipole-pair excitations in an elastic medium and are sparsely distributed \cite{hasyim2021theory,chacko2021elastoplasticity} (\cref{fig:figure_3}(c)-(i)).
In contrast, we find that above $T_\mathrm{o}$ the liquid is able to relax upon the first IS jump through the formation of free dipolar excitations (\cref{fig:figure_3}(c)-(ii)).
Thus, $T_\mathrm{o}$ signals a change in relaxation mechanism between the supercooled and high-temperature regimes. 

\begin{figure*}[t]
    \centering
    \hspace{0.08in}\includegraphics[width=0.95\textwidth]{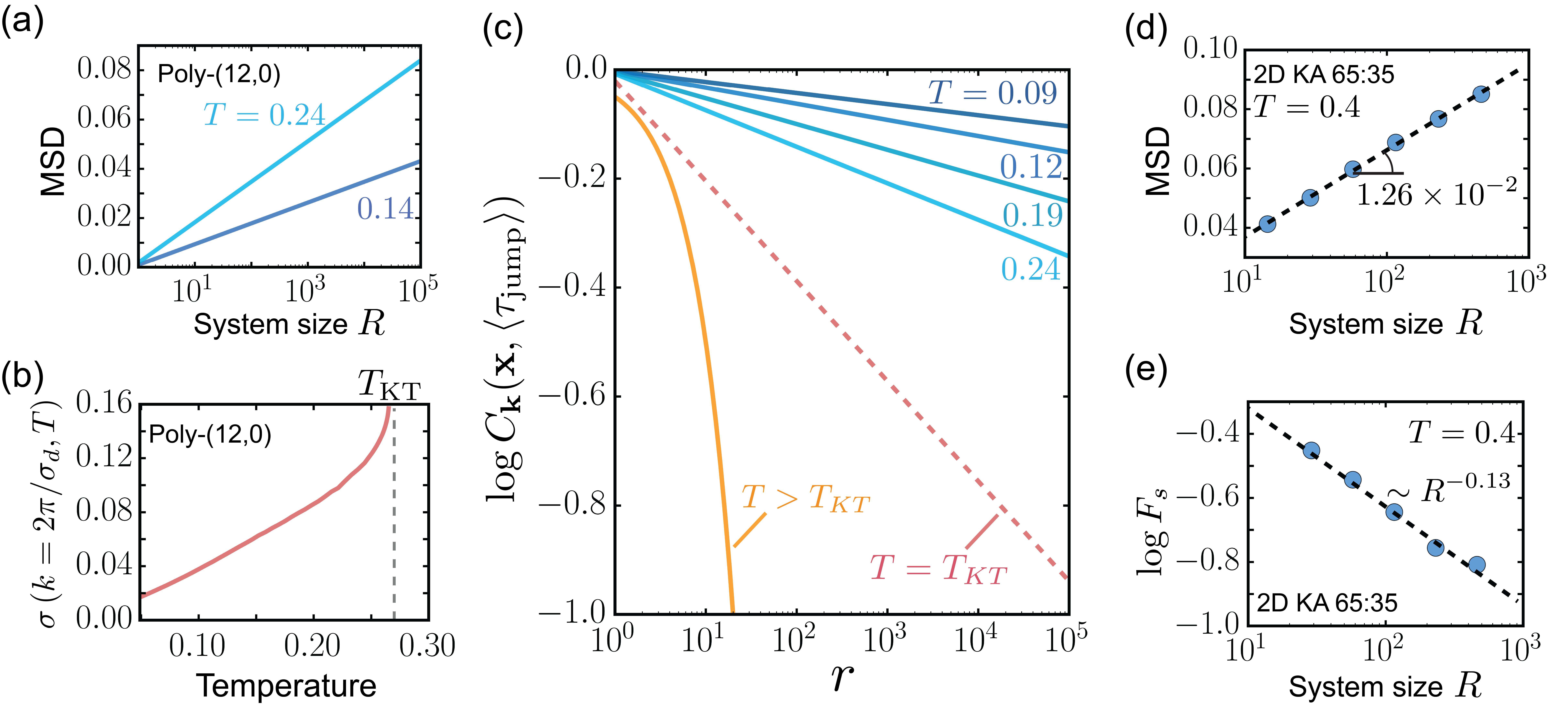}
    \caption{(a) Predicted mean square displacement (MSD) vs. system size $R$ for the Poly-(12,0) model at two different temperatures below $T_\mathrm{KT}$. In two dimensions, the elastic nature of the supercooled liquid in the glassy regime leads to a logarithmic system-size dependence. (b) The power-law exponent $\sigma(k,T)$ in \cref{eq:sigma_exponent} vs. temperature for the Poly-(12,0) model for $k = 2 \pi$. At low temperatures, it increases linearly as a result of $Y^\mathrm{R} \rightarrow Y^\mathrm{IS}$, while for $T\rightarrow T_\mathrm{KT}$ it increases abruptly because the material loses its rigidity. (d) Predicted spatial dependence of the correlation function $C_{\* k}(\* x, \langle \tau_\mathrm{jump} \rangle)$ for the Poly-(12,0) model glass former. (c) Fitted  MSD vs. $R$ data for the two-dimensional (2D) Kob-Andersen (KA) 65:35 model at $T=0.4$ and $t=100$~\cite{shiba2019local}. (e) Fitted $F_s\left(k,t\right)$ vs. $R$ data for the 2D KA 65:35 model at $T=0.4$~\cite{shiba2019local}, where $k=2 \pi/\sigma_\mathrm{d}$, $\sigma_\mathrm{d}=1$, and $t=100$. Note that the scalings for the Poly-(12,0) model are predictions, and are yet to be tested extensively through large scale molecular simulations. }\label{fig:figure_4}
\end{figure*}

Our theory also provides a way to study displacement and density correlations of supercooled liquids at the intermediate timescales. As the theory suggests that supercooled liquid behaves as a solid at timescales $t \simeq \langle \tau_\mathrm{jump} \rangle$, we can use an effective Gaussian field theory~\cite{kardar2007statistical} for a fluctuating elastic medium with renormalized elastic constants~\cite{nelson1978study,nelson1979dislocation} to find that the MSD is (SM, Sec.~5.5)
\begin{equation}
 \left\langle |\mathbf{u} (\langle \tau_\mathrm{jump} \rangle)|^2 \right\rangle  
     \simeq k_\mathrm{B} T \frac{(3-\nu^\mathrm{R})(1+\nu^\mathrm{R})}{2 \pi  Y^\mathrm{R}} \ln \frac{R}{\xi^*} \,, \label{eq:ujumpfinal}   
\end{equation}
where $\xi^*$ sets the smallest lengthscale for which the elastic Gaussian field theory is valid. The lengthscale $\xi^*$ is also connected to the characteristic size of the bound-dipole pairs, and becomes $\xi^* \sim \mathcal{O}(a_\mathrm{dpl}) \sim \mathcal{O}(\sigma_\mathrm{d})$ as $T \to 0$, where $\sigma_\mathrm{d}$ is the particle diameter (SM, Sec.~5.5). \Cref{eq:ujumpfinal} is a signature of the Mermin-Wagner fluctuations in a 2D solid, and is consistent with recent observations from experiments and computer simulations \cite{flenner2015fundamental,shiba2016unveiling,illing2017mermin,vivek2017long,tarjus2017glass}. 
In \cref{fig:figure_4}(a), we also plot the logarithmic scaling, as predicted for the Poly-(12,0) model glass former.

The signatures of Mermin-Wagner fluctuations can also be found in the self-part of the intermediate scattering function $F_s(k, t )$, where the theory suggests (SM, Sec 5.5)   
\begin{subequations}
    \begin{alignat}{2}
    F_s(k,\langle \tau_\mathrm{jump}\rangle ) 
     & \simeq \left(\frac{R}{\xi^*}\right)^{-\frac{\sigma\left(k,T\right)}{2}} \label{eq:fskt_correlation} , 
     \\
    \sigma\left(k,T\right) & = k_\mathrm{B} T \frac{k^2\left(3-\nu^\mathrm{R}\right)\left(1+\nu^\mathrm{R}\right)}{4 \pi Y^\mathrm{R}} \,,  \label{eq:sigma_exponent}
    \end{alignat}
\end{subequations}
which is valid for wavenumber $k \in [0, 2\pi/\xi^*]$. Since $\xi^* \sim \mathcal{O}(\sigma_\mathrm{d})$ as $T \to 0$, \cref{eq:fskt_correlation} becomes valid at lengthscales in which we typically measure relaxation dynamics, e.g., $k=2 \pi/\sigma_\mathrm{d}$. Thus, the theory suggests that relaxation, as measured by $F_s(k,t)$, proceeds faster with increasing system size due to Mermin-Wagner fluctuations alone. 

In addition to finite-size effects, Mermin-Wagner fluctuations can also be probed spatially. To this end, we introduce an order-parameter field $\rho_{\* k}(\* x,t) = e^{i \* k \cdot \* u(\* x,t) }$ based on the displacement field $\* u(\* x,t)$, which is computed from the particle displacement $\* u^\alpha(t)$. The spatial correlation function of $\rho_{\* k}(\* x,t)$ at $t \simeq \langle \tau_\mathrm{jump} \rangle$ can be written as
\begin{align}
C_{\* k}(\* x, \langle \tau_\mathrm{jump} \rangle) &:= 
\left\langle\rho_\mathbf{k}\left(\mathbf{x},\langle \tau_\mathrm{jump} \rangle \right)\rho_\mathbf{-k}\left(\mathbf{0},0 \right)\right\rangle 
\\
&\simeq \left(\frac{\left\lvert\mathbf{x}\right\rvert}{R}\right)^{-\sigma\left(k,T\right)} \,, \label{eq:rhok_rhok_correlation} 
\end{align}
see SM, Sec 5.5.
\Cref{eq:rhok_rhok_correlation} implies that spatial fluctuations of the order-parameter field with respect to an initial inherent state exhibit power-law correlations at intermediate timescales, thereby indicating quasi-long range order reminiscent of the 2D crystalline phases \cite{kosterlitz1972long,kosterlitz1973ordering}. 
Such power-law decay is in contrast to the exponential decay found in past studies of spatial correlations in supercooled liquids \cite{hallett2018local,berthier2011dynamical}, where structural order parameters were used to probe static correlations without reference to an initial inherent state. 
The power-law exponent $\sigma(k,T)$, which also enters into the finite-size scaling in \cref{eq:fskt_correlation}, increases with higher temperature, as seen in \cref{fig:figure_4}(b) for the predicted $\sigma(k,T)$ of the Poly-(12,0) model glass former. This results in a faster decay of correlations as  $T \to T_\mathrm{KT}$ (\cref{fig:figure_4}(c)), with an expected exponential decay above the onset temperature corresponding to the fluid phase.

While Figs.~\ref{fig:figure_4}(a)-(c) show predicted finite-size scalings for the Poly-(12,0) model glass former that are yet to be tested, we validate such scalings to available literature data. For instance, the 2D Kob-Andersen (KA) model \cite{kob1995testing} has been studied in large-scale molecular simulations \cite{shiba2019local}, with data available for both the MSD and $F_s(k,t)$ at the same temperature for various system sizes.  
If the theory is applicable to the 2D KA model, then the data for MSD and $F_s(k,t)$ must not only follow the expected finite size scalings in  \cref{eq:ujumpfinal,,eq:fskt_correlation}, but the corresponding exponents must also be related to each other by a factor of $k^2/4$.
Indeed, Figs.~\ref{fig:figure_4}(d) and \ref{fig:figure_4}(e) show the logarithmic and power-law finite-size scalings of the MSD and $F_s(k,t)$, respectively, for the 2D KA model when $t\approx 100$ and $k=2\pi/\sigma_\mathrm{d}$. Note that these scalings also hold for a range of intermediate time scales less than the relaxation times. 
Furthermore, the fitted slope for $F_s(k,t)$ is  $\sigma(k=2\pi/\sigma_\mathrm{d},T)/2 \approx 0.13$,
which quantitatively agrees with the one obtained from MSD, which is $(1.26 \times 10^{-2}) k^2 /4\approx 0.124$ where $\sigma_\mathrm{d}=1$ \cite{shiba2019local}. 
These results constitute a first step in validating the theory in terms of the consequences for finite-size effects at the level of MSD and density autocorrelations. Further tests for temperature dependence of the exponents corresponding to the finite-size effects are left for future work.

$\\$\emph{Conclusions and Discussion}.--- 
In summary, we construct a theory for the onset temperature $T_\mathrm{o}$ of glassy dynamics in two-dimensions (2D), starting from the isoconfigurational ensemble \cite{widmer2004reproducible} as a basis for the statistical mechanics of excitations in supercooled liquids. 
The resulting framework allows us to derive the Arrhenius form of the rate/concentration of excitations, $c_\mathrm{eq}(T) \sim e^{-\beta \tilde{E}_\mathrm{c}}$, which is empirically found in DF theory when performing rate calculations from molecular simulations \cite{Keys2011,hasyim2021theory}. 
To understand the onset of glassy dynamics, excitations are represented as interacting geometric dipoles, a description unique to the 2D nature of the liquids. 
In parallel to the KTHNY theory, $T_\mathrm{o}$ can be described as a binding-unbinding transition of dipolar elastic excitations as well as melting of inherent states, and the predicted $T_\mathrm{o}$ is in reasonable agreement across six different model glass formers (\cref{tab:predictions_KT}). 
The theory also enables studies on displacement and density correlations, where the predicted finite-size scalings are consistent with recent observations of Mermin-Wagner fluctuations from simulations and experiments in 2D glass formers \cite{flenner2015fundamental,shiba2016unveiling,illing2017mermin,vivek2017long,tarjus2017glass}.

Since the origin of $T_\mathrm{o}$ lies within the isoconfigurational ensemble, the inherent-state melting transition is a hidden transition that is not directly observable from the liquid thermodynamic properties. 
Furthermore, analyzing this transition is difficult upon noting the required separation of timescales (\cref{eq:inttimescale}), which may limit the range of applicability of our theory near $T_\mathrm{o}$. 
However, the reasonable agreement between the predicted and observed $T_\mathrm{o}$ suggests that the theory is useful in interpreting the emergence of glassy dynamics at $T_\mathrm{o}$ as the onset of inherent-state stability against excitation fluctuations. 
Further tests need to be performed through simulations using the isoconfigurational ensemble~\cite{widmer2004reproducible}, and in particular on the temperature-dependence of the MSD and spatial correlations in density fluctuations. 
Experiments through (quasi)-2D colloidal systems~\cite{illing2017mermin,vivek2017long} also provide an additional platform to test the theory through finite-size effects. 
Lastly, even though our work sheds light on the onset temperature in 2D, the nature of the onset temperature in three dimensions (3D) remains an open question. 
The corresponding theory for 3D may require an extension of the geometric-charges framework to 3D, in a way that yields an analogous inherent-state melting scenario.
We leave the possibility of such a theory for future work.

$\\$\emph{Acknowledgements}.--- The authors acknowledge Cory Hargus for insightful comments on the manuscript. M.R.H. and K.K.M. were entirely supported by the Director, Office of Science, Office of Basic Energy Sciences, of the U.S. Department of Energy under Contract No. DEAC02-05CH1123. D.F. (dfrag) acknowledges support from the Miller Institute for Basic Research in
Science at University of California, Berkeley. 

\bibliographystyle{apsrev4-1}
\bibliography{refs}

\end{document}